\documentclass[11pt]{article}

\usepackage[final]{acl}

\usepackage{times}
\usepackage{latexsym}

\usepackage[T1]{fontenc}

\usepackage[utf8]{inputenc}

\usepackage{microtype}

\usepackage{inconsolata}

\usepackage{graphicx}

\usepackage{amsmath}
\usepackage{amssymb}

\usepackage{booktabs}
\usepackage{array}

\usepackage{float}

\usepackage{placeins}

\title{Debiasing as a Measurement Intervention: Calibrated Ties and Resolution Loss in LLM-as-a-Judge Evaluation}

\author{Liang Zhao\textsuperscript{*} \quad
  Yong Wang\textsuperscript{*\textdagger} \quad
  Jiangzhe Chen\textsuperscript{*} \\
  Shanghai University of International Business and Economics \\
  \texttt{\{25349212,25349213,25349216\}@suibe.edu.cn} \\
  {\small \textsuperscript{*}Equal contribution. \textsuperscript{\textdagger}Corresponding author: \texttt{25349213@suibe.edu.cn}.}}

\begin{document}
\maketitle

\begin{abstract}
LLM-as-a-judge protocols are commonly debiased by instructing judges to ignore presentation cues such as citation formatting, source labels, and evidence-display style. We show that this intervention can suppress bias while damaging the resolution of the measurement instrument. We introduce TraceJudgeBench, a diagnostic benchmark for auditing citation-like artifacts in RAG and agent-workflow evaluation, covering content-equivalent pairs, citation ablations, correctness conflicts, human-validated soft and moderate quality gaps, prompt-strength ladders, decoupled judging, and a controlled workflow-ranking probe. Across GPT-5.5, Claude Sonnet 4.6, and DeepSeek V4-Flash, stronger anti-citation prompts reduce worse-cited wins from up to 50.5\% to 0\%; yet some operating points already convert validated moderate-gap decisions into Tie before the strict stress-test endpoint, while correctness-conflict accuracy remains at or above 93.0\%. A second, 50-pair FinQA moderate-gap construction reproduces the qualitative frontier, and open-weight Qwen2.5-14B-Instruct-AWQ and Gemma-3-12B-IT runs reproduce the central HotpotQA frontier. TRACE-style decoupling recovers 96.5--100.0\% better-plain resolution across the reported settings. Human validation separates three meanings of Tie: correct equivalence Tie, calibrated soft-boundary Tie, and resolution-destroying Tie on validated quality gaps. We frame debiasing as a measurement intervention whose bias suppression, resolution retention, Tie cost, and protocol cost must be reported jointly. The supplementary artifact contains benchmark splits, prompts, raw judge outputs, validation summaries, and analysis scripts.
\end{abstract}

\section{Introduction}

LLM-as-a-judge evaluation has become a standard measurement layer for open-ended generation, instruction following, retrieval-augmented generation (RAG), agent workflows, and arena-style model comparison. Its appeal is clear: LLM judges are scalable, inexpensive relative to human annotation, and applicable to outputs whose quality cannot be captured by exact match or lexical overlap. However, this scalability also makes the judge itself part of the measurement system. When a judge is sensitive to presentation, order, verbosity, model identity, or output format, the resulting score no longer reflects only the underlying quality of the candidates.

This problem becomes especially sharp in RAG and agent-workflow evaluation. Modern systems rarely produce answer-only text; they typically emit source titles, citation markers, sentence identifiers, retrieved snippets, tool traces, verification statements, and other evidence-display artifacts \citep{bragg-etal-2026-astabench, chong-etal-2026-talk, molfese-etal-2026-retraceqa}. These artifacts have a dual status. In some tasks, they are superficial wrappers that should not influence the judgment when the substantive answer is unchanged. In others, they are part of the evaluated behavior, because traceability, support-chain completeness, and evidence use are legitimate dimensions of quality. A judge protocol must therefore decide whether citation-like artifacts are irrelevant presentation cues or useful evidence of task performance.

A common remedy is to debias the judge with a stricter rubric: instruct it to ignore citation formatting, source labels, evidence-display style, length, and polish, and to focus only on correctness or substantive content. This intervention is often warranted. If two candidates are content-equivalent and differ only in citation-looking wrappers, a strict judge should return Tie. In this setting, Tie is not indecision; it is the desired suppression of a spurious preference.

Debiasing, however, is not merely bias removal. It is a measurement intervention that can change what the benchmark is able to distinguish. Many evaluation settings require more than separating clearly correct from clearly wrong answers. They also require distinguishing candidates that share the same final answer but differ in support-chain completeness, reasoning specificity, evidence coverage, or workflow faithfulness. In this region, a strict anti-presentation rubric may remove useful quality resolution together with superficial presentation sensitivity. The hidden cost is not necessarily a collapse in clear correctness. Instead, the measurement signal can shift from biased preference to abstention: a judge that once over-rewarded citation-like display may become unable to select the better answer when the quality gap is real but moderate.

We study this trade-off through the lens of calibrated Tie versus resolution-destroying Tie. Tie is not a single phenomenon. On content-equivalent pairs, Tie is correct. On soft-boundary pairs where human annotators also disagree or mostly choose Tie, Tie may be a calibrated expression of uncertainty. But on human-validated moderate-quality gaps, Tie is costly: the protocol has lost benchmark resolution. This distinction is essential because debiasing methods that look successful under equivalence tests may still fail as measurement tools for quality-sensitive evaluation.

We introduce TraceJudgeBench, a diagnostic benchmark for auditing whether LLM-as-a-judge protocols preserve decision resolution after suppressing citation-like presentation cues. We run the full diagnostic matrix on three commercial judges: GPT-5.5, Claude Sonnet 4.6, and DeepSeek V4-Flash. We additionally run the central moderate-gap prompt-frontier and TRACE-recovery audits on two open-weight judges. The open-weight runs do not cover every split, but they reproduce the main protocol-level pattern: stricter anti-presentation instructions can trade superficial-win suppression for Tie inflation, while quality-preserving decoupling recovers moderate-gap resolution.

\begin{figure*}[!t]
  \centering
  \includegraphics[width=0.85\textwidth]{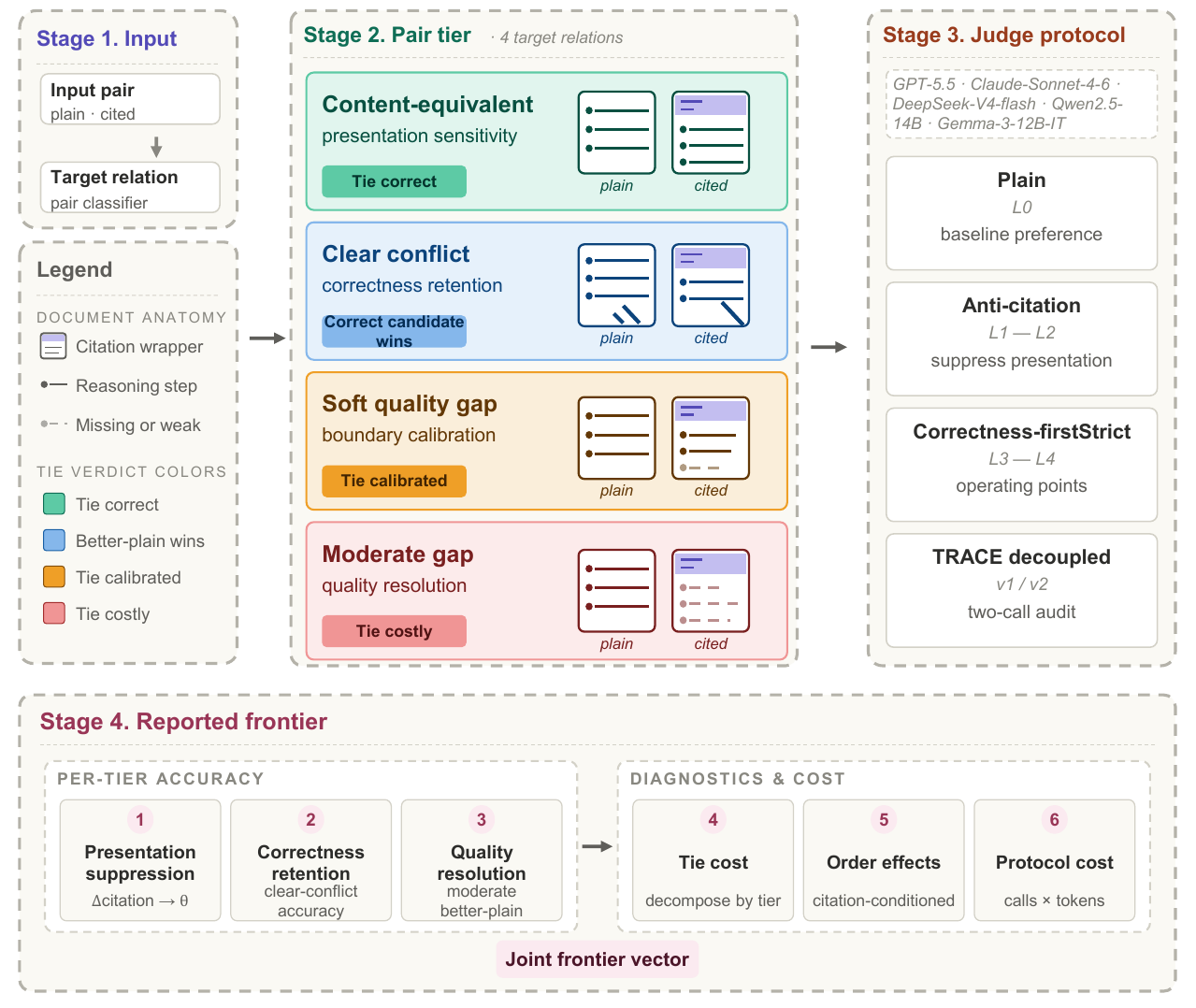}
  \caption{TraceJudgeBench protocol-audit framework}
  \label{fig:pipeline}
\end{figure*}

Figure~\ref{fig:pipeline} summarizes TraceJudgeBench as a protocol-audit framework: input pairs are organized by target relation, evaluated under prompt and protocol families, and reported through a bias--resolution frontier rather than a single aggregate score.

Our contributions are:

\begin{enumerate}
  \item We frame debiasing in LLM-as-a-judge evaluation as a measurement intervention, not merely as bias removal.
  \item We introduce TraceJudgeBench, a diagnostic benchmark for auditing citation-like artifacts and debiasing side effects in RAG and agent-workflow judging.
  \item We identify debiasing-induced resolution loss, in which stricter anti-presentation rubrics suppress citation-like wins but convert human-validated moderate quality gaps into Tie.
  \item We distinguish calibrated Tie from resolution-destroying Tie using content-equivalent, soft-boundary, and moderate-gap human validation tiers.
  \item We show that the central bias--resolution frontier appears across commercial and open-weight judges, while reporting the full diagnostic matrix for the three commercial judges.
\end{enumerate}

\section{Related Work}

\paragraph{LLM-as-a-judge evaluation and bias.}

Pairwise LLM judging is widely used for open-ended model evaluation, preference comparison, reward-model assessment, and arena-style ranking. MT-Bench and Chatbot Arena established LLM judges as a scalable alternative to human preference collection, while documenting limitations such as position bias, verbosity bias, self-enhancement, and constrained reasoning \citep{zheng-etal-2023-judging}. AlpacaEval and Length-Controlled AlpacaEval show that automatic evaluators can reward superficial correlates of quality, especially length, unless the protocol is explicitly debiased \citep{dubois-etal-2024-length}. RewardBench broadens this line by evaluating reward models and LLM judges across heterogeneous preference settings \citep{lambert-etal-2025-rewardbench}. Other work treats the judge itself as an object of evaluation, studying position, verbosity, self-enhancement, output-format bias, hidden preference leakage, and automatically discovered judge biases \citep{wang-etal-2024-large, long-etal-2025-llms, lai-etal-2026-biasscope, li-etal-2026-preference}. Our work builds on this literature but asks a different question: after a bias is mitigated, does the protocol still preserve the resolution needed for quality-sensitive evaluation?

\paragraph{Evidence display and RAG evaluation.}

RAG evaluation has made evidence use a central evaluation target. Frameworks such as RAGAS and ARES evaluate faithfulness, answer relevance, and context quality, reflecting the fact that answer correctness alone does not fully capture grounded-generation quality \citep{es-etal-2024-ragas, saad-falcon-etal-2024-ares}. Attribution work makes a related point: answer correctness and citation faithfulness are not equivalent, since an answer can be correct while its cited evidence is incomplete, irrelevant, or unsupported \citep{wallat-etal-2025-correctness}. Citation-like artifacts therefore have a dual status. When the target is only final-answer correctness, source labels or citation formatting may be confounders. When the target includes traceability, evidence coverage, or auditability, the same artifacts may carry task-relevant information. We adopt a controlled-comparison approach in the spirit of work that holds content constant while varying representation \citep{zhang-etal-2026-same}: hold or validate substantive quality while varying citation-like presentation, then test whether debiasing preserves useful resolution.

\paragraph{Resolution, abstention, and benchmark separability.}

Reliable evaluation requires not only aggregate accuracy, but also the ability to distinguish meaningful differences. Work on micro-benchmarking, item-response theory, and rank-aware evaluation shows that benchmarks can preserve aggregate trends while failing to separate close systems or saturated item regions \citep{yauney-etal-2026-how, zhou-etal-2026-lost, hu-etal-2026-eip}. Tie and abstention are also not inherently failures: abstention can be desirable when inputs are ambiguous or genuinely uncertain \citep{zhai-etal-2026-abstain, noh-etal-2026-multi}. TraceJudgeBench connects these points at the protocol level. We ask whether a debiasing prompt suppresses spurious presentation preference while preserving validated quality gaps, and we interpret Tie relative to the target relation of each pair rather than treating all Tie as either good calibration or bad indecision.

\section{Problem Setup and Metrics}

We study pairwise LLM-as-a-judge evaluation. Given a task \(x\) and two candidate outputs \(y_A\) and \(y_B\), a judge protocol \(P\) returns one of three labels: \(A\), \(B\), or Tie. Each comparison is run under an ordered presentation of the two candidates, so that position effects can be measured rather than assumed away.

Our focus is on pairs in which one output carries a citation-like evidence-display artifact, such as source titles, citation markers, sentence identifiers, retrieved snippets, evidence blocks, or verification-looking statements. We call this output the \textbf{cited candidate}, and the other output the \textbf{plain} or \textbf{base candidate}. The artifact may be purely presentational, as in content-equivalent pairs, or it may coexist with real differences in support-chain completeness, reasoning specificity, or evidence coverage.

We evaluate judge protocols under three target relations. In \textbf{content-equivalent pairs}, the two candidates have the same substantive answer and differ only in citation-like presentation; the desired outcome is Tie. In \textbf{clear correctness conflicts}, one candidate is correct and the other is wrong; the desired outcome is to select the correct candidate regardless of citation display. In \textbf{graded quality conflicts}, both candidates share the correct final answer but differ in support-chain completeness or reasoning specificity; these pairs test whether a judge preserves resolution on real but moderate quality gaps.

On equivalence or near-equivalence pairs, we report the cited-candidate win rate \(W_{\text{cited}}\), plain/base win rate \(W_{\text{plain}}\), and Tie rate \(T\). We also report the signed presentation effect:

\[
\Delta_{\text{citation}} = W_{\text{cited}} - W_{\text{plain}}.
\]

A positive value means citation-like display raises the probability of winning; a value near zero with high Tie indicates successful suppression of presentation preference. On correctness-conflict pairs, we report correct-candidate wins, wrong-candidate wins, and Tie. On moderate graded pairs, we report better-plain wins, worse-cited wins, and Tie, and summarize directional resolution with:

\[
\Delta_{\text{quality}} = W_{\text{better-plain}} - W_{\text{worse-cited}}.
\]

Tie is interpreted by tier. Equivalence Tie measures bias suppression, soft-boundary Tie may indicate calibration, and moderate-gap Tie measures lost resolution. We also report candidate-order gaps and protocol cost. Single-call judge protocols require one judge call per ordered comparison; TRACE-style decoupled judging requires two calls, one normalization call and one masked judging call. For signed effects and order gaps, we use pair-level bootstrap confidence intervals with 2,000 resamples; for proportions near 0\% or 100\%, we use Wilson score intervals. The end-to-end audit flow---pair construction, target-relation tagging, protocol selection, and frontier reporting---is summarized in Figure~\ref{fig:pipeline}.

\section{TraceJudgeBench}

TraceJudgeBench is a protocol-audit benchmark, not a universal RAG leaderboard. Its goal is to test whether LLM-as-a-judge protocols suppress citation-like presentation sensitivity while preserving the resolution needed for quality-sensitive evaluation. Each split isolates one measurement question, such as presentation sensitivity, naturalistic validity, artifact triggers, correctness retention, quality resolution, order effects, or protocol cost.

Table~\ref{tab:components} lists the seven diagnostic components, their sizes, target relations, and judge coverage. Citation-like artifacts include source titles, citation markers, sentence identifiers, retrieved snippets, evidence blocks, and verification-looking statements. In some splits, they are superficial wrappers; in others, they coexist with real support-chain differences and test whether a judge can separate presentation from evidence quality.

\begin{table*}[!t]
  \centering
  \small
  \begin{tabular}{>{\raggedright\arraybackslash}p{2.2cm}>{\raggedright\arraybackslash}p{2.4cm}>{\raggedright\arraybackslash}p{2.0cm}>{\raggedright\arraybackslash}p{3.0cm}>{\raggedright\arraybackslash}p{2.8cm}}
    \toprule
    \textbf{Component} & \textbf{Size} & \textbf{Target relation} & \textbf{Measurement question} & \textbf{Judge coverage} \\
    \midrule
    Mechanical equivalence & 200 HotpotQA + 200 FinQA pairs & content-equivalent & Does citation-like display create spurious preference? & commercial full matrix \\
    Natural citation-looking pairs & 1,000 HotpotQA pairs; 200 human-validated & near-equivalent & Does the effect persist in more natural outputs? & commercial plain full; strict on 200-pair subset \\
    Citation ablations & six variants & content-controlled & Which artifact triggers the preference? & commercial full matrix \\
    Position controls & 1,200 Claude + 400 DeepSeek comparisons & identical candidates & Is the effect generic position bias or citation-conditioned order coupling? & targeted commercial controls \\
    Clear and reverse conflicts & 150 pairs, run across prompt paraphrases and orders & correctness conflict & Does citation display override obvious correctness? & commercial plain/strict prompt paraphrases; decoupled single prompt \\
    Soft and moderate graded conflicts & 100 soft + 100 HotpotQA moderate + 50 FinQA moderate pairs & quality conflict & Does debiasing preserve validated quality gaps? & commercial full; open-weight Qwen/Gemma on HotpotQA moderate \\
    Workflow-ranking probe & 200 HotpotQA queries, four workflow variants & system comparison & Can presentation sensitivity affect controlled tournament scores? & commercial + open-weight controlled probe, not full leaderboard \\
    \bottomrule
  \end{tabular}
  \caption{TraceJudgeBench diagnostic components}
  \label{tab:components}
\end{table*}

We construct 200 HotpotQA \citep{yang-etal-2018-hotpotqa} mechanical citation pairs and 200 FinQA \citep{chen-etal-2021-finqa} mechanical citation pairs. In each pair, the cited candidate preserves the same substantive content as the plain candidate but adds citation-like presentation, such as source labels or evidence-display blocks. Any systematic preference for the cited candidate therefore indicates presentation sensitivity rather than improved answer quality. To complement these artificial but causally controlled examples, we add 1,000 natural HotpotQA citation-looking pairs whose final answers match and whose outputs differ naturally in source labeling or evidence display.

Mechanical pairs are further expanded into six ablation variants: full citation, empty shell, source-only, evidence-only, sentence-id-only, and length-control. These variants separate source-like authority, raw evidence text, formatting shell, sentence identifiers, and added length. If judges merely prefer longer answers, the length-control variant should win; if they respond to source-like authority or traceability display, the full-citation and source-only variants should be strongest.

Correctness-conflict probes test whether citation-like display overwhelms obvious correctness. In each clear conflict, the plain candidate keeps the correct answer, while the cited candidate receives citation-like presentation but has its final answer deliberately changed to an incorrect one. Reverse conflicts swap the direction, making the cited candidate correct and the plain candidate wrong. We also add position-only controls with identical plain/no-citation candidates to separate generic position bias from citation-conditioned order coupling.

The central resolution test uses two 100-pair HotpotQA graded tiers. In the \textbf{moderate tier}, both candidates give the same correct final answer, but the plain candidate provides a complete evidence or reasoning chain while the cited candidate omits or weakens one necessary supporting hop, entity link, or evidence-chain step. In the \textbf{soft tier}, both candidates match in final answer and main evidence, but the plain candidate is somewhat more specific while the cited candidate is more generic. The soft tier is intentionally near the human decision boundary. A supplementary 50-pair FinQA moderate tier tests a different mechanism: both candidates share the numeric answer and evidence, but the cited-style candidate omits an intermediate denominator, base, or conversion step.

Two independent annotators validate the natural and graded samples. Table~\ref{tab:human-validation} summarizes the validation signals. The HotpotQA moderate tier is the main human-validated basis for the resolution-loss claim: both annotators label 91/100 pairs as better-plain, and raw agreement is 100.0\%. For FinQA, two blind annotators agree on final-answer equivalence for 50/50 pairs and on reasoning completeness and preference for 49/50; a third independent annotator adjudicates the disputed item. By contrast, only 1/100 soft pairs are labeled better-plain by both original annotators, confirming that soft is a boundary tier rather than a stable superiority tier. We report PABAK and Gwet's AC1 alongside raw agreement instead of Cohen's kappa, which is known to collapse on prevalence-skewed agreement \citep{cicchetti-feinstein-1990-high}.

\begin{table*}[!t]
  \centering
  \small
  \begin{tabular}{lrrrr>{\raggedright\arraybackslash}p{3.8cm}}
    \toprule
    \textbf{Validation set} & \textbf{$n$} & \textbf{Raw agreement} & \textbf{PABAK} & \textbf{Gwet's AC1} & \textbf{Main human-label signal} \\
    \midrule
    Natural equivalence & 200 & 92.0\% & 0.840 & 0.918 & 183/200 both equivalent \\
    Graded soft & 100 & 96.0\% & 0.920 & 0.958 & 1/100 both better plain \\
    Graded moderate & 100 & 100.0\% & 1.000 & 1.000 & 91/100 both better plain \\
    Graded overall & 200 & 98.0\% & 0.960 & 0.976 & moderate is resolution-loss evidence \\
    \bottomrule
  \end{tabular}
  \caption{Human validation summary}
  \label{tab:human-validation}
\end{table*}

We compare single-call judge protocols along a prompt-strength ladder: L0 plain preference, L1 anti-citation warning, L2 quality-ignore-citation, L3 correctness-first, and L4 strict tie-required. L4 is treated as a stress-test endpoint rather than a recommended deployment rubric; the main frontier is interpreted from the full L0--L4 trajectory, especially the L2/L3 operating points that resemble common debiasing instructions. We also test TRACE-style decoupled judging. TRACE first normalizes away citation/source presentation while preserving final answer, support-chain completeness, reasoning specificity, unsupported claims, omitted support, and partial-support signals; it then performs masked pairwise judging over the normalized records. TRACE is not a new model or a deployment-ready judge. It is a two-call audit baseline that makes presentation suppression and quality-resolution retention more auditable.

We evaluate the full matrix on GPT-5.5, Claude Sonnet 4.6, and DeepSeek V4-Flash, using temperature 0 and JSON-mode outputs where supported. We additionally run the central moderate-gap prompt frontier, TRACE recovery, and controlled ranking probes on two open-weight judges, Qwen2.5-14B-Instruct-AWQ and Gemma-3-12B-IT (full identifier and serving details in Appendix~\ref{sec:appendix-d}). The open-weight experiments test whether the main frontier generalizes, but they are not presented as a full open-weight audit.

\section{Results}

We organize the results around presentation sensitivity, correctness retention, quality resolution, open-weight replication, and TRACE recovery.

\subsection{Citation-Like Display Creates Spurious Preference}

Under plain judging, citation-like presentation can substantially affect pairwise decisions even when substantive content is unchanged. On mechanical equivalence pairs, GPT-5.5 selects the cited candidate 85.8\% of the time on HotpotQA, with Claude and DeepSeek at 50.9\% and 47.2\% (Figure~\ref{fig:equivalence}). The presentation effect also appears in the soft graded tier: GPT-5.5 selects the worse-cited candidate in 50.5\% of cases under plain judging---a preference that anti-citation prompts later suppress to 0\%. Natural citation-looking pairs show the same phenomenon under less artificial conditions: source labels, evidence blocks, and verification-looking statements can be treated as signals of support even when the underlying answer is equivalent or near-equivalent.

Citation ablations show that the trigger is not simply length. Length-control and evidence-only variants do not account for the full effect. Instead, full-citation and source-only variants are strongest, suggesting that judges respond to source-like authority or traceability display rather than to additional tokens alone. Rationale analysis (Appendix~\ref{sec:appendix-g}, Table~\ref{tab:keyword-rates}) supports this interpretation: citation-win rationales frequently describe the cited candidate as more supported, verifiable, or traceable, with source/citation keyword rates near 100\% on citation-win decisions across all three commercial judges.

Candidate order also matters: position-only controls separate generic order bias from citation-conditioned order coupling (Appendix Table~\ref{tab:position-controls}).

\subsection{Strict Rubrics Suppress Presentation Bias While Retaining Clear-Conflict Accuracy}

Stronger anti-citation prompts successfully suppress presentation preference in equivalence settings. As the prompt ladder moves from L0 plain preference to L4 strict tie-required judging, worse-cited wins fall to 0\% in the commercial judges. This is the intended outcome on content-equivalent pairs: the judge should not reward citation-looking wrappers when the substantive answer is unchanged.

Importantly, this suppression does not imply a general failure to judge correctness. On clear and reverse correctness conflicts, commercial judges mostly retain high accuracy under stricter prompts. Clear-conflict probes remain at or above 93.0\% across the reported commercial settings; full strict-prompt correctness-family results are included in the supplementary artifact. This separates two failure modes. The problem is not that debiasing makes judges unable to recognize obvious wrong answers. Rather, the risk appears in a harder region: pairs where both candidates share the correct final answer but differ in support-chain completeness or reasoning specificity.

\begin{figure*}[!t]
  \centering
  \includegraphics[width=0.78\textwidth]{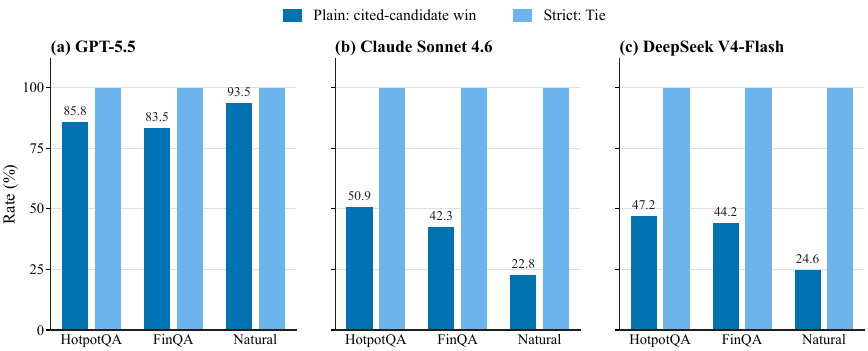}
  \caption{Equivalence suppression}
  \label{fig:equivalence}
\end{figure*}

Figure~\ref{fig:equivalence} visualizes this first-stage success: strong anti-citation prompting suppresses high cited-candidate win rates and moves the decision mass toward Tie. If the audit stopped here, the debiasing intervention would appear successful.

\subsection{Moderate Quality Gaps Reveal Resolution Loss}

The moderate graded tier exposes the hidden cost of strict debiasing. In this tier, both candidates share the correct final answer, but the plain candidate has a complete support chain while the cited candidate omits or weakens a necessary supporting step. Human validation establishes that this is a real quality gap rather than a formatting preference.

For abstention-heavy judges, resolution loss begins before the strict stress-test endpoint. At L3, Claude's better-plain rate falls to 71.5\% and DeepSeek's to 79.0\%, while Tie rises to 28.5\% and 21.0\%, respectively; Gemma-3-12B-IT shows an even sharper L3 Tie jump in the open-weight replication. L4 exposes the endpoint behavior: Claude and DeepSeek drop to 11.0--15.0\% better-plain decisions, and GPT-5.5 falls to 2.0\%. Worse-cited wins remain near zero, so the failure is not a return of citation preference. It is resolution loss.

The soft tier explains why Tie cannot be interpreted uniformly. In soft cases, human any-Tie prevalence is 96\%; at L2/L3, commercial-judge Tie rates are 96--100\%, with item-level raw agreement of 0.94--0.96. This supports, but does not by itself prove, the calibrated-Tie interpretation. The moderate contrast is more informative: human Tie prevalence is 9\%, whereas Claude and DeepSeek assign Tie to 42\% and 33\% of L3 items, with Tie precision of 0.119 and 0.152. Appendix Table~\ref{tab:soft-frontier} gives the aggregate ladder. This tiered evidence is why TraceJudgeBench reports Tie by target relation rather than only aggregate Tie rates.

\subsection{The Bias--Resolution Frontier Appears Across Commercial and Open-Weight Judges}

\begin{figure*}[!t]
  \centering
  \includegraphics[width=0.94\textwidth]{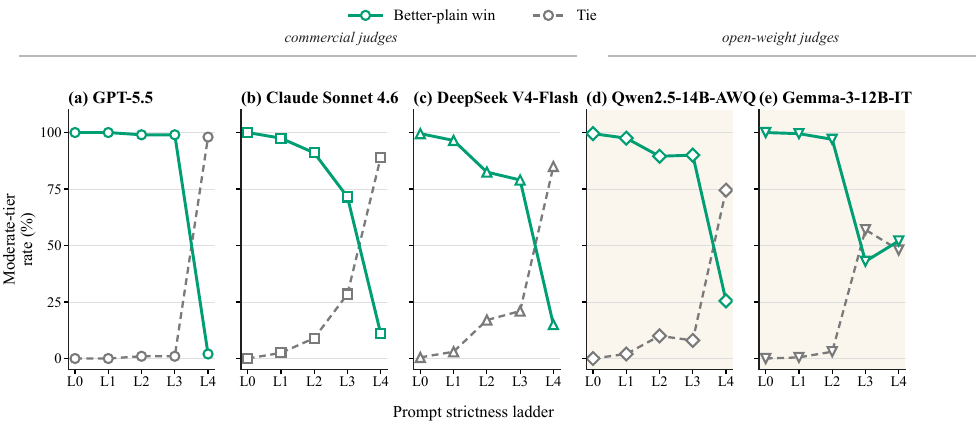}
  \caption{Bias-resolution frontier over prompt strictness}
  \label{fig:frontier}
\end{figure*}

Figure~\ref{fig:frontier} summarizes the central trade-off. Moving along the prompt ladder can suppress worse-cited wins, but for some judges it also shifts decision mass into Tie and reduces better-plain resolution on moderate gaps. We therefore do not rely on L4 alone: the frontier is the trajectory across L1--L3 operating points, the L4 stress test, and TRACE-style decoupling. Paired tests confirm the endpoint changes: for every commercial judge, the HotpotQA L0-to-L4 loss and L4-to-TRACE recovery are significant (exact McNemar, $p<0.001$). For example, Claude loses 89.0 points of better-plain resolution (95\% paired-bootstrap CI [84.5, 93.0]) and GPT-5.5 loses 98.0 points (CI [96.0, 99.5]). A protocol improves the frontier only when it suppresses spurious citation preference without destroying validated quality resolution.

The open-weight judges are not run on the full diagnostic matrix, but they reproduce the central frontier. Qwen2.5-14B-Instruct-AWQ retains more moderate-gap resolution under strict prompting than the most abstention-heavy commercial judges, while Gemma-3-12B-IT shows a sharp Tie jump at the L2$\rightarrow$L3 transition, although its L4 endpoint partially recovers better-plain decisions. These differences matter: the frontier is not a single universal curve. However, both open-weight judges support the main protocol-level claim that debiasing can trade presentation suppression for resolution loss.

\subsection{TRACE-Style Decoupling Recovers Moderate-Gap Resolution}

TRACE-style decoupling shows that resolution loss is not inevitable. The correctness-only decoupled variant removes presentation artifacts but can still discard quality-relevant support-chain structure. The quality-preserving variant, \texttt{decoupled\_quality\_v2}, explicitly preserves final answer, support-chain completeness, reasoning specificity, unsupported claims, omitted support, and partial-support signals before masked judging.

This distinction matters. On the moderate tier, \texttt{decoupled\_quality\_v2} recovers or preserves better-plain resolution across all five judges included in the frontier analysis. Commercial judges recover to 97.0--100.0\% better-plain decisions, and the two open-weight judges recover 96.5--98.5\% better-plain resolution. Worse-cited wins remain at 0\%. Thus, the moderate-gap failure is not an unavoidable consequence of ignoring citation-like presentation; it depends on whether the protocol preserves quality-relevant evidence structure while removing display cues. The full L0--L4 ladder appears in Appendix Table~\ref{tab:moderate-frontier}, with TRACE per-judge details in Table~\ref{tab:trace-recovery}.

The independent FinQA construction shows the same qualitative boundary. Across 50 pairs, five prompt levels, and both candidate orders, L0--L2 better-plain rates are 95--97\% for GPT-5.5, 92--98\% for Claude, and 88--95\% for DeepSeek. At L4, Tie reaches 91\%, 100\%, and 99\%, respectively; TRACE recovers 98/100, 100/100, and 100/100 ordered decisions with no Ties. DeepSeek already falls to 46\% better-plain at L3, so the result is not confined to the definitional L4 endpoint or to HotpotQA support-hop omission.

\subsection{Controlled Ranking Shows System-Level Risk}

Finally, we test whether presentation sensitivity can affect controlled system comparison. We construct a workflow-ranking probe over 200 HotpotQA queries and four workflow variants. The goal is not to produce a new RAG leaderboard, but to test whether citation-like display can alter tournament-style rankings when substantive differences are controlled.

The ranking probe confirms the system-level risk. Citation-like display can improve a variant's apparent standing under ordinary judging, while stricter or decoupled protocols reduce this advantage (full per-protocol scores and direct full-vs-full comparisons in Appendix~\ref{sec:appendix-e8}--\ref{sec:appendix-e8b}, Tables~\ref{tab:workflow-scores} and~\ref{tab:direct-comparisons}). This result connects the pairwise diagnostics to practical evaluation: if presentation artifacts influence pairwise wins, they can also affect aggregate rankings, model selection, and reported progress.

\section{Discussion}

Our results suggest that debiasing an LLM judge should be treated as a measurement intervention rather than a purely corrective prompt edit. A stricter instruction can remove an unwanted presentation preference, but it can also change the operating point of the evaluator. The key question is therefore not only whether a protocol reduces bias, but what kind of measurement signal remains after the bias is removed.

First, bias suppression is not sufficient. A protocol that suppresses citation preference on content-equivalent pairs may look successful under a narrow bias test, but still fail as a measurement instrument if it converts human-validated moderate differences into Tie. Clear-conflict accuracy alone is also insufficient as a safety check. A judge can identify obvious correct-versus-wrong cases while failing to distinguish moderate support-chain differences.

Second, Tie has multiple meanings. On content-equivalent pairs, Tie is a success because it suppresses a spurious preference. On soft-boundary pairs, Tie may be calibrated because human annotators also do not reliably prefer one candidate. On human-validated moderate gaps, however, Tie is a cost: it means the protocol failed to preserve a distinction that humans judged to be real. Debiasing studies should therefore report Tie by tier rather than only aggregate win rates.

Third, citation-like artifacts should be treated neither as automatically valid evidence nor as automatically irrelevant formatting. If the evaluation target is final-answer correctness only, citation-like display should usually be controlled or ignored. If the target includes traceability, evidence use, or workflow faithfulness, the protocol must preserve those quality dimensions while preventing source-like presentation from acting as a shortcut.

TRACE demonstrates that the failure mode is protocol-dependent. It requires approximately 3.7--4.3$\times$ the total tokens and 5.3--5.9$\times$ the dollar cost of a one-call judge (about USD 18.6--21.6 per 1,000 comparisons under the reported GPT-5.5 pricing). It may also introduce extraction errors, so it should be understood as an audit protocol rather than a universal deployment solution.

\section{Conclusion}

We introduced TraceJudgeBench, a diagnostic benchmark for auditing how LLM-as-a-judge protocols handle citation-like artifacts in RAG and agent-workflow evaluation. Our results show that stronger anti-citation instructions can successfully suppress spurious citation preference on content-equivalent pairs, but this success can hide resolution loss on human-validated moderate quality gaps. Tie is therefore not a single outcome: it is correct on equivalence pairs, potentially calibrated on soft-boundary pairs, and costly on validated moderate gaps. TRACE-style decoupling recovers much of the lost resolution by removing citation-like presentation while preserving quality-relevant support structure. Overall, debiasing should be reported as a frontier over presentation suppression, correctness retention, quality resolution, Tie cost, order effects, and protocol cost.

\section*{Limitations}

TraceJudgeBench is a diagnostic benchmark, not a universal RAG or agent leaderboard. Its controlled and partly template-built splits isolate measurement questions, which improves interpretability but does not cover the full diversity of naturally occurring quality differences, retrieval pipelines, agent tools, or user-facing domains \citep{seo-etal-2026-simuhome, shu-etal-2026-dare, dong-etal-2025-finch}. Surface regularities may also make the tampered side detectable without full task reasoning; ablations and order controls reduce but do not eliminate this concern. The full diagnostic matrix is evaluated on three commercial judges; open-weight experiments focus on the central HotpotQA moderate frontier, TRACE recovery, and ranking probes rather than every split. Most human validation uses two annotators; the FinQA disagreement receives independent adjudication, but broader multi-annotator validation would further strengthen the benchmark. Finally, TRACE is costly, may introduce extraction errors, and its fixed-field normalization explicitly records missing support. Stage 1 can therefore perform part of the quality assessment before masked comparison, so high recovery alone does not isolate the contribution of Stage 2.

\section*{Ethics Statement}

LLM-as-a-judge systems are increasingly used to compare models, tune systems, select checkpoints, and report benchmark progress. If judge protocols reward superficial presentation artifacts, they may overstate systems that produce source-looking outputs without better evidence use. If debiasing protocols collapse moderate quality differences into Tie, they may understate genuine improvements in traceability or support-chain completeness. TraceJudgeBench is intended for diagnostic auditing, not for ranking people or making high-stakes decisions. The annotation task concerns model outputs rather than personal or sensitive user data. Annotators were recruited from graduate students familiar with NLP evaluation, worked under blind candidate ordering, and were compensated at an hourly rate consistent with local minimum-wage requirements. HotpotQA-derived material is distributed under CC BY-SA 4.0 and FinQA under the MIT license; released derivatives retain source identifiers and license notices. Details of recruitment, workload, and adjudication are provided in Appendix B.

\section*{Acknowledgments}

We thank the anonymous reviewers and area chairs for their constructive feedback. We also thank the annotators who contributed to the human-validation study.

\bibliography{custom}

\newpage
\appendix
\setcounter{table}{0}
\setcounter{figure}{0}
\renewcommand{\thetable}{A\arabic{table}}
\renewcommand{\thefigure}{A\arabic{figure}}
\section*{Appendix}
\label{sec:appendix}

This appendix provides construction details, prompt templates, extended results, and reproducibility notes for TraceJudgeBench. The main paper reports only the compact diagnostic matrix and the central bias-resolution frontier; the appendix expands the audit so that each result can be traced back to a split, protocol, and model setting.

\FloatBarrier
\textbf{Cross-reference to the main paper.} The table below maps each main-paper float to the appendix material that supports it.

\begin{table*}[!htbp]
  \centering
  \small
  \begin{tabular}{>{\raggedright\arraybackslash}p{3.6cm}>{\raggedright\arraybackslash}p{5.8cm}>{\raggedright\arraybackslash}p{4.0cm}}
    \toprule
    \textbf{Main paper} & \textbf{Appendix material} & \textbf{Topic} \\
    \midrule
    Figure~\ref{fig:pipeline} (audit pipeline) & Benchmark construction details (Appendix \ref{sec:appendix-a}) & Per-split and per-protocol design \\
    Table~\ref{tab:components} (diagnostic components) & Benchmark construction details (Appendix \ref{sec:appendix-a}) & Per-split design \\
    Table~\ref{tab:human-validation} (human validation) & Table \ref{tab:appendix-validation}; kappa note (\ref{sec:appendix-b3}) & Inter-annotator reliability \\
    \S5.5 compact frontier + TRACE & Tables \ref{tab:moderate-frontier} and \ref{tab:trace-recovery} & Per-judge full ladder and recovery \\
    Figure~\ref{fig:equivalence} (equivalence suppression) & Tables \ref{tab:equivalence-summary} and \ref{tab:equivalence-suppression} & Per-judge $\times$ per-split breakdown \\
    Figure~\ref{fig:frontier} (bias--resolution frontier) & Tables \ref{tab:moderate-frontier}, \ref{tab:trace-recovery}, \ref{tab:soft-frontier} & Moderate and soft prompt-ladder behavior \\
    \S5.1 rationale analysis & Table \ref{tab:keyword-rates} & Source/citation keyword rates \\
    \S5.6 controlled ranking & Tables \ref{tab:workflow-scores} and \ref{tab:direct-comparisons} & Workflow probe details \\
    \bottomrule
  \end{tabular}
  \caption{Cross-reference from each main-paper float to the supporting
  appendix material.}
  \label{tab:appendix-map}
\end{table*}

\section{Benchmark Construction Details}
\label{sec:appendix-a}

This section supports the \emph{Mechanical equivalence}, \emph{Natural citation-looking pairs}, \emph{Citation ablations}, \emph{Clear and reverse conflicts}, and \emph{Soft and moderate graded conflicts} rows of main-paper Table~\ref{tab:components}. The audit pipeline is shown in main-paper Figure~\ref{fig:pipeline}; this appendix provides the construction details behind that diagram.

\subsection{Mechanical Equivalence Pairs}
\label{sec:appendix-a1}

Mechanical equivalence pairs isolate presentation sensitivity under maximal content control. For each source item we construct two candidates:

\begin{itemize}
  \item \textbf{Plain candidate.} Contains the task-relevant answer, evidence, reasoning, and final response without citation-like wrappers.
  \item \textbf{Cited candidate.} Preserves the same substantive answer content but adds citation-like evidence-display artifacts: source titles, citation markers, sentence identifiers, retrieved snippets, or verification-looking blocks.
\end{itemize}

The cited candidate introduces no new task-relevant facts. A systematic cited-candidate win on this split therefore reflects presentation sensitivity rather than improved answer quality. We construct 200 HotpotQA mechanical pairs and 200 FinQA mechanical pairs to test whether the effect appears in both open-domain multi-hop QA and financial/numerical reasoning settings.

\subsection{Natural Citation-Looking Pairs}
\label{sec:appendix-a2}

Mechanical pairs provide causal control but can look artificial. We therefore add 1,000 natural HotpotQA citation-looking pairs. These pairs preserve final-answer agreement but allow naturally occurring differences in source labels, evidence display, or verification-like wording. A 200-pair subset receives human validation.

The natural split is used in two ways. The full 1,000-pair set supports the plain-judge presentation-sensitivity audit. The 200-pair validated subset supports stricter protocol comparisons and human-equivalence checks. Thus, the natural split is not a full strict-protocol matrix over all 1,000 pairs.

\subsection{Citation Ablation Variants}
\label{sec:appendix-a3}

To identify which artifact types trigger presentation sensitivity, we derive six ablation variants from the HotpotQA mechanical pairs.

\begin{table*}[!t]
  \centering
  \small
  \begin{tabular}{p{2.5cm}p{4.5cm}p{5.5cm}}
    \toprule
    \textbf{Variant} & \textbf{Description} & \textbf{Diagnostic purpose} \\
    \midrule
    Full citation & Source labels, evidence display, and citation-like formatting & Tests the complete artifact bundle \\
    Empty shell & Citation-like structure without informative evidence content & Tests formatting shell effects \\
    Source-only & Source titles or source labels without full evidence blocks & Tests source-like authority cues \\
    Evidence-only & Evidence text without source/citation presentation & Tests raw evidence content \\
    Sentence-id-only & Sentence or passage identifiers without source display & Tests identifier artifacts \\
    Length-control & Added non-substantive length on the perturbed/cited side without citation structure & Tests length as an alternative explanation \\
    \bottomrule
  \end{tabular}
  \caption{Citation ablation variants.}
  \label{tab:ablation-variants}
\end{table*}

If length alone explained the effect, the length-control variant would dominate. If raw evidence explained the effect, the evidence-only variant would dominate. If source-like authority or traceability display explains the effect, full-citation and source-only variants should be strongest. Appendix~\ref{sec:appendix-e1} reports the resulting rates.

\subsection{Correctness Conflicts}
\label{sec:appendix-a4}

Clear-conflict probes test whether citation-like display overwhelms obvious correctness. Each clear conflict contains one correct candidate and one incorrect candidate. In the main direction, the plain candidate is correct and the cited candidate is wrong; reverse conflicts swap the roles.

The split contains 150 pairs sampled from three domains---open-domain multi-hop QA, financial reasoning, and short code-style tasks---run across three prompt paraphrases and both candidate orders. Decoupled correctness judging uses a single prompt version and both orders; it is therefore a targeted probe rather than a full prompt-paraphrase matrix.

\subsection{Graded Soft and Moderate Quality Conflicts}
\label{sec:appendix-a5}

The graded splits test quality resolution when both candidates share the same correct final answer.

In the \textbf{moderate tier}, the plain candidate provides a complete evidence or reasoning chain. The cited candidate includes citation-like display but omits or weakens one necessary supporting hop, entity link, or evidence-chain step. The target relation is better-plain.

In the \textbf{soft tier}, both candidates match in final answer and main evidence. The plain candidate is somewhat more specific in reasoning, while the cited candidate is more generic. The target relation is intentionally near the human decision boundary.

Construction proceeds in three steps. (1) Lock the shared final answer and supporting facts from locally checked HotpotQA traces. (2) Generate either a moderate support-chain omission or a soft specificity reduction using fixed templates. (3) Validate the resulting pairs with blind human annotation (Appendix~\ref{sec:appendix-b}).

The FinQA robustness split uses a different moderate-gap construction. Both candidates retain the same final numeric answer and the same financial evidence, while the cited-style candidate omits an intermediate denominator, base quantity, or conversion step from the arithmetic chain. The split contains 50 pairs and is evaluated at five prompt levels in both candidate orders.

\subsection{Representative Examples}
\label{sec:appendix-a6}

These examples illustrate the construction patterns of Appendices~\ref{sec:appendix-a1}, \ref{sec:appendix-a4}, and \ref{sec:appendix-a5}.

\textbf{Mechanical equivalence example.} \emph{Task:} Were Scott Derrickson and Ed Wood of the same nationality? The plain candidate gives the final answer ``yes'' using evidence that Scott Derrickson is American and Ed Wood was an American filmmaker. The cited candidate preserves the same answer and evidence, then appends a citation block reusing the same source sentences. The target relation is content-equivalent; the gold relation is Tie.

\textbf{Clear correctness conflict example.} For the same task, the plain candidate keeps the correct answer ``yes''. The cited candidate displays source-like citation markers over the same evidence but changes the final answer to ``no''. The target relation is correctness conflict; the gold relation is better-plain.

\textbf{Moderate quality-gap example.} \emph{Task:} Where did the descendants of the Black Seminoles settle? The plain candidate answers ``Coahuila, Mexico'' and includes both the Mascogos evidence and the Black Seminoles bridge. The cited candidate gives the same final answer and citation-like display but includes only the Mascogos evidence, omitting part of the support chain. The target relation is better-plain; a Tie on this item type is treated as resolution loss.

\textbf{FinQA numeric-gap example pattern.} Both candidates report the same percentage and cite the same financial values. The plain candidate shows the required subtraction and division by the correct base, whereas the cited-style candidate jumps from the inputs to the final percentage without exposing the denominator or conversion step. The target relation is better-plain.

\textbf{Soft boundary example.} In the soft tier, both candidates share the same final answer and main evidence, but the plain candidate gives a somewhat more explicit reasoning sentence while the cited candidate gives a more generic support statement. Human validation shows that these items sit near the decision boundary; Tie is therefore interpreted as possible calibration rather than automatic failure.

\section{Human Validation}
\label{sec:appendix-b}

\subsection{Annotation Tasks}
\label{sec:appendix-b1}

Annotators see the task and two candidate outputs without knowing which side carries citation-like display or which label is expected. For natural equivalence validation, annotators use one primary label.

\begin{table*}[!t]
  \centering
  \small
  \begin{tabular}{>{\raggedright\arraybackslash}p{3.4cm}>{\raggedright\arraybackslash}p{8.6cm}}
    \toprule
    \textbf{Label} & \textbf{Meaning} \\
    \midrule
    \texttt{equivalent} & Both candidates give the same final answer and are equally correct for the task. Citation/source formatting introduces no correctness-relevant difference. \\
    \texttt{base\_more\_correct} & The base/plain candidate is more correct, more faithful to evidence, or less misleading. \\
    \texttt{citation\_more\_correct} & The citation-looking candidate is more correct, more faithful to evidence, or less misleading. \\
    \texttt{unclear} & The item cannot be confidently judged from the provided traces. \\
    \bottomrule
  \end{tabular}
  \caption{Human annotation labels for natural citation-looking validation.}
  \label{tab:annotation-labels}
\end{table*}

Annotators additionally check whether the citation-looking candidate introduces new factual content, whether the citation/source display could affect presentation preference without changing correctness, and whether any mismatch exists in final answer, numeric answer, named entity, date, or yes/no polarity.

For graded soft and moderate validation, annotators answer: (i) whether the final answers are equivalent; (ii) whether one candidate has more complete support; (iii) whether one candidate has more specific reasoning; (iv) whether citation/source display adds substantive information; and (v) the final preference label---Candidate A, Candidate B, or Tie.

The final preference label is interpreted relative to the target relation of each pair. A Tie on equivalence pairs supports the desired equivalence relation. A Tie on soft-boundary pairs may be calibrated. A Tie on moderate pairs is treated as resolution loss when human consensus establishes a better-plain direction. Natural-pair validation is reported separately from mechanical minimal-pair validation; the two are not merged in main-paper rates.

\paragraph{Scope of human validation.}
The original validation uses two independent annotators recruited from graduate students; both were fluent English speakers familiar with NLP evaluation tasks. Each labeled 400 pairs, consisting of 200 natural-equivalence pairs and 200 graded quality-gap pairs. The workload was estimated at approximately 8 hours per annotator, and annotators were compensated at an hourly rate consistent with local minimum-wage requirements. We chose blind pairwise annotation because the target relation is local to a pair rather than an absolute quality score. The HotpotQA moderate-tier claim relies on high-consensus items: 91/100 pairs are labeled better-plain by both annotators, with 100.0\% raw agreement over the tier. Two additional blind annotators label the 50 FinQA pairs; they agree on final-answer equivalence for 50/50 pairs and on reasoning completeness and preference for 49/50. A third independent annotator adjudicates the single disputed item. We do not claim that this panel is sufficient for a universal preference benchmark. The released workbook includes guidelines, item identifiers, blind candidate order, per-item labels, and adjudication flags.

\subsection{Validation Summary}
\label{sec:appendix-b2}

We repeat the main-paper validation table here for completeness, then add the kappa note that does not fit in the main 8 pages.

\begin{table*}[!t]
  \centering
  \small
  \begin{tabular}{lrrrr>{\raggedright\arraybackslash}p{4.5cm}}
    \toprule
    \textbf{Validation set} & \textbf{$n$} & \textbf{Raw agreement} & \textbf{PABAK} & \textbf{Gwet's AC1} & \textbf{Main human-label signal} \\
    \midrule
    Natural equivalence & 200 & 92.0\% & 0.840 & 0.918 & 183/200 both equivalent \\
    Graded soft & 100 & 96.0\% & 0.920 & 0.958 & 1/100 both better plain \\
    Graded moderate & 100 & 100.0\% & 1.000 & 1.000 & 91/100 both better plain \\
    FinQA moderate & 50 & 98.0\% & --- & --- & 49/50 initial preference agreement; one adjudication \\
    Graded overall & 200 & 98.0\% & 0.960 & 0.976 & moderate is resolution-loss evidence \\
    \bottomrule
  \end{tabular}
  \caption{Human validation summary (repeated from main-paper Table~2 for self-contained discussion of reliability metrics).}
  \label{tab:appendix-validation}
\end{table*}

\subsection{Why Cohen's Kappa Is Not Reported}
\label{sec:appendix-b3}

The main-paper table omits Cohen's kappa because these validation sets are intentionally prevalence-skewed. In the natural-equivalence set, most examples are expected to be equivalent; in the graded-moderate set, most items are expected to be better-plain. Under such marginal skew, kappa is known to collapse even when raw agreement is high---the ``high agreement but low kappa'' paradox documented by Cicchetti and Feinstein (1990). We therefore report raw agreement, PABAK, and Gwet's AC1 as the primary reliability signals. Per-item kappa values can be released as supplementary data if needed; they do not change the qualitative conclusion that annotator consensus is strong on the moderate tier.

\section{Judge Protocols}
\label{sec:appendix-c}

\subsection{Single-Call Prompt Ladder}
\label{sec:appendix-c1}

TraceJudgeBench uses a prompt-strength ladder to study the operating point of a judge protocol. The ladder is not intended as an exhaustive prompt search; it provides interpretable settings from ordinary preference judging to strict anti-presentation judging.

\begin{table*}[!t]
  \centering
  \small
  \begin{tabular}{>{\raggedright\arraybackslash}p{1.2cm}>{\raggedright\arraybackslash}p{3cm}>{\raggedright\arraybackslash}p{7cm}}
    \toprule
    \textbf{Level} & \textbf{Name} & \textbf{Intended behavior} \\
    \midrule
    L0 & Plain preference & Choose the better overall answer, allowing Tie \\
    L1 & Anti-citation warning & Warn that citation-like display may be superficial \\
    L2 & Quality-ignore-citation & Evaluate answer quality while ignoring citation formatting unless it adds substantive information \\
    L3 & Correctness-first & Prioritize correctness and substantive support over presentation, length, or polish \\
    L4 & Strict tie-required & Return Tie when candidates are equivalent in final-answer correctness, even if one appears more cited or polished \\
    \bottomrule
  \end{tabular}
  \caption{Single-call prompt ladder.}
  \label{tab:prompt-ladder}
\end{table*}

Representative instruction excerpts for the moderate-tier ladder are shown in Table~\ref{tab:prompt-excerpts}.

\begin{table*}[!t]
  \centering
  \small
  \begin{tabular}{>{\raggedright\arraybackslash}p{1.2cm}>{\raggedright\arraybackslash}p{10cm}}
    \toprule
    \textbf{Level} & \textbf{Prompt excerpt} \\
    \midrule
    L0 & Choose which candidate response is better overall for the task. If both candidates are equally good, choose Tie. \\
    L1 & Choose which candidate response is better overall, but do not prefer a response merely because it looks cited, sourced, or workflow-like. \\
    L2 & Choose the response that is better overall, but ignore citation formatting, source labels, evidence-block style, response length, and polish by themselves. Consider those features only if they change task correctness or genuinely necessary information. \\
    L3 & First judge task correctness. If both answers are equally correct, use genuinely task-relevant support completeness or reasoning specificity as a tiebreaker, but do not use citation-like formatting, source titles, evidence display style, length, or polish as advantages by themselves. \\
    L4 & Prefer a candidate only when it is more correct for the task. If both candidates give the same correct final answer and differ only in citation formatting, source labels, evidence display, length, style, verification wording, support-chain detail, or reasoning specificity, choose Tie. \\
    \bottomrule
  \end{tabular}
  \caption{Moderate-tier prompt excerpts. Full prompt templates are released with the artifact.}
  \label{tab:prompt-excerpts}
\end{table*}

All single-call prompts request JSON with \texttt{winner} in \texttt{\{A, B, Tie\}} and a short rationale.

\paragraph{Interpreting L4.}
L4 is a deliberately strict stress-test endpoint. It is useful because it reveals how far a judge can move toward abstention when presentation cues are aggressively discounted, but it is not treated as the only or recommended debiasing protocol. The main paper therefore interprets the full ladder, with special attention to L2 and L3: these levels resemble common evaluation instructions that ask judges to ignore formatting, polish, citation display, length, and non-substantive presentation while still selecting the better answer when substantive support differs. Resolution loss at L2/L3 is therefore more informative than the L4 endpoint alone.

\paragraph{FinQA prompt-paraphrase check.}
We test wording sensitivity with Claude on 50 FinQA pairs in both candidate orders. Two faithful L2 paraphrases yield 93--95\% better-plain and 0--2\% Tie, close to the original L2 result (92\%). A faithful L3 paraphrase yields 91\% better-plain and 0\% Tie, in the same direction as the original L3 result (94\%). A boundary variant that explicitly requires Tie whenever final answers match produces 100\% Tie. We label it L4-like rather than a faithful L3 paraphrase. This check indicates that the direction is stable for the tested faithful paraphrases while confirming that whether reasoning completeness may break a tie is the critical instruction boundary.

\subsection{Generic Pairwise Judge Output Schema}
\label{sec:appendix-c2}

All protocols request a structured label:

\begin{verbatim}
{
  "winner": "A | B | Tie",
  "reason": "brief explanation"
}
\end{verbatim}

When structured output is not natively enforced by the endpoint, responses are parsed with a conservative label extractor. Ambiguous or unparsable outputs are logged for manual inspection rather than silently coerced.

\subsection{TRACE-Style Decoupled Audit}
\label{sec:appendix-c3}

TRACE-style decoupling uses two calls per ordered comparison in our implementation:

\begin{enumerate}
  \item \textbf{Joint normalization call.} Normalize both candidates in one call, removing or masking citation/source presentation while preserving final answer, support-chain completeness, reasoning specificity, unsupported claims, omitted support, and partial-support signals.
  \item \textbf{Masked judging call.} Compare the normalized records without seeing the original citation-like display.
\end{enumerate}

The quality-preserving version used for the main frontier is \texttt{decoupled\_quality\_v2}. The correctness-only version, \texttt{decoupled\_v1}, is useful as a diagnostic contrast because it can suppress presentation cues while failing to preserve quality-relevant support-chain structure.

The Stage 1 normalization prompt instructs the judge to remove citation formatting, source labels, source titles, bracketed references, appended citation blocks, evidence-display style, surface polish, and wording differences that do not alter task-relevant content. It explicitly preserves final answer, task-relevant factual claims, reasoning needed to connect claims to the final answer, support-chain completeness, explanation specificity, missing support, and whether the response merely states an answer without support.

To reduce the risk that TRACE merely moves the cue into another representation, the normalized record uses fixed fields rather than free-form rewritten answers: final answer, support facts, reasoning links, missing or weakened support, unsupported claims, and residual uncertainty. Source titles, bracketed citations, document labels, passage identifiers, and source-count features are masked or omitted. The masked judging call sees only these normalized fields and the task, not the original candidate text or the original side labels. Released logs include both the original pair and normalized record so that reviewers can audit whether presentation cues were removed and whether quality-relevant support was preserved.

The Stage 2 masked comparison prompt receives only the normalized records. It judges task correctness, support-chain completeness, and reasoning specificity. It instructs the judge not to infer quality from citations, source labels, formatting, length, polish, or evidence-display style, and to choose Tie when the normalized content is task-quality equivalent.

TRACE is an audit protocol, not a deployment-ready solution. Its role is to test whether resolution loss is inevitable under debiasing; the answer is no, provided the protocol preserves support-chain quality during normalization.

Stage 1 is not assumed to be a neutral formatter. Because its fixed schema records missing or weakened support, normalization can perform part of the quality assessment before Stage 2. The released original-to-normalized records make this behavior auditable, but the high recovery rate alone cannot distinguish sound decomposition from decisions already encoded by Stage 1. Isolating these contributions is left to future ablations.

Relative to a one-call judge, TRACE uses approximately 3.7--4.3$\times$ total tokens and 5.3--5.9$\times$ dollar cost in the reported accounting, corresponding to about USD 18.6--21.6 per 1,000 comparisons under GPT-5.5 standard pricing at the time of the experiments.

\section{Model and Serving Details}
\label{sec:appendix-d}

\subsection{Commercial Judges}
\label{sec:appendix-d1}

The full diagnostic matrix is evaluated on three commercial judges.

\begin{table*}[!t]
  \centering
  \small
  \begin{tabular}{>{\raggedright\arraybackslash}p{2.8cm}>{\raggedright\arraybackslash}p{4cm}>{\raggedright\arraybackslash}p{4cm}>{\raggedright\arraybackslash}p{2.8cm}}
    \toprule
    \textbf{Judge} & \textbf{Identifier used in logs} & \textbf{Additional endpoint metadata} & \textbf{Coverage} \\
    \midrule
    GPT-5.5 & \texttt{gpt-5.5} & endpoint-reported snapshot \texttt{gpt-5.5-2026-04-23} & full diagnostic matrix \\
    Claude Sonnet 4.6 & \texttt{claude-sonnet-4-6} & provider snapshot not fully exposed by the endpoint & full diagnostic matrix \\
    DeepSeek & \texttt{deepseek-v4-flash} & endpoint-reported alias used for all DeepSeek runs & full diagnostic matrix \\
    \bottomrule
  \end{tabular}
  \caption{Commercial judge identifiers and coverage.}
  \label{tab:commercial-judges}
\end{table*}

All runs use temperature 0. Structured JSON output is used where supported. Provider-side model snapshots may not be fully exposed by all endpoints; released logs include call dates, endpoint aliases, endpoint-reported snapshots when available, response metadata, prompt templates, retry rules, and raw A/B/Tie decisions.

\subsection{Open-Weight Judges}
\label{sec:appendix-d2}

The open-weight experiments focus on robustness replication of the central moderate-gap frontier rather than full TraceJudgeBench coverage.

\begin{table*}[!t]
  \centering
  \small
  \begin{tabular}{>{\raggedright\arraybackslash}p{3.2cm}>{\raggedright\arraybackslash}p{4cm}>{\raggedright\arraybackslash}p{3.5cm}>{\raggedright\arraybackslash}p{2.8cm}}
    \toprule
    \textbf{Judge} & \textbf{Checkpoint / serving identifier} & \textbf{Coverage} & \textbf{Role} \\
    \midrule
    Qwen2.5-14B-Instruct-AWQ & \texttt{Qwen2.5-14B-Instruct-AWQ} & full moderate prompt frontier, TRACE recovery, controlled ranking probe & open-weight robustness replication \\
    Gemma-3-12B-IT & \texttt{Gemma-3-12B-IT} & full moderate prompt frontier, TRACE recovery, controlled ranking probe & open-weight robustness replication \\
    \bottomrule
  \end{tabular}
  \caption{Open-weight judge identifiers and coverage.}
  \label{tab:openweight-judges}
\end{table*}

Both open-weight judges are served via a local OpenAI-compatible endpoint. Released logs include the exact \texttt{MODEL\_NAME} returned by the endpoint, base URL type, quantization/runtime, call dates, and inference parameters. The main paper does not claim full open-weight coverage of equivalence, ablation, correctness-conflict, position-control, or paraphrase-audit splits.

\section{Extended Results}
\label{sec:appendix-e}

\subsection{Equivalence and Natural Protocol Summary}
\label{sec:appendix-e1}

This expanded view supports main-paper Figure~\ref{fig:equivalence} (equivalence suppression) and the first claim of \S5.1.

\begin{table*}[!t]
  \centering
  \small
  \begin{tabular}{>{\raggedright\arraybackslash}p{2.8cm}>{\raggedright\arraybackslash}p{2.5cm}rrrrr}
    \toprule
    \textbf{Model} & \textbf{Protocol} & \textbf{$n$} & \textbf{Tie} & \textbf{Citation win} & \textbf{Base win} & \textbf{Signed effect} \\
    \midrule
    GPT-5.5 & Mechanical Plain        & 1200 & 14.2 & 85.8 & 0.0 & 85.8 \\
    GPT-5.5 & Mechanical Strict       & 1200 & 100.0 & 0.0 & 0.0 & 0.0 \\
    GPT-5.5 & Decoupled Mechanical    & 400  & 100.0 & 0.0 & 0.0 & 0.0 \\
    GPT-5.5 & Natural Plain           & 6000 & 6.4  & 93.5 & 0.1 & 93.4 \\
    GPT-5.5 & Natural Strict          & 1200 & 99.9 & 0.1  & 0.0 & 0.1 \\
    Claude Sonnet 4.6 & Mechanical Plain     & 1200 & 48.8 & 50.9 & 0.2 & 50.7 \\
    Claude Sonnet 4.6 & Mechanical Strict    & 1200 & 100.0 & 0.0 & 0.0 & 0.0 \\
    Claude Sonnet 4.6 & Decoupled Mechanical & 400  & 100.0 & 0.0 & 0.0 & 0.0 \\
    Claude Sonnet 4.6 & Natural Plain        & 6000 & 77.1 & 22.8 & 0.1 & 22.7 \\
    Claude Sonnet 4.6 & Natural Strict       & 1200 & 100.0 & 0.0 & 0.0 & 0.0 \\
    DeepSeek & Mechanical Plain         & 1200 & 52.5 & 47.2 & 0.3 & 46.8 \\
    DeepSeek & Mechanical Strict        & 1200 & 100.0 & 0.0 & 0.0 & 0.0 \\
    DeepSeek & Decoupled Mechanical     & 400  & 100.0 & 0.0 & 0.0 & 0.0 \\
    DeepSeek & Natural Plain            & 6000 & 74.4 & 24.6 & 1.0 & 23.7 \\
    DeepSeek & Natural Strict           & 1200 & 99.8 & 0.0  & 0.2 & -0.2 \\
    \bottomrule
  \end{tabular}
  \caption{Equivalence and natural protocol summary.}
  \label{tab:equivalence-summary}
\end{table*}

Here \(n\) is the number of judge rows after applying the protocol-specific pair subset, prompt paraphrases, and candidate-order settings, not simply the number of source questions. Plain and strict single-call rows include prompt paraphrases and candidate-order presentation; decoupled rows use a smaller audited subset and one protocol setting. The released manifest records the exact pair subset, prompt family, and order expansion for each row count.

This table supports the first-stage claim in the main paper: stricter and decoupled protocols suppress citation wins on equivalence-like pairs, mostly by moving decisions to Tie.

\subsection{Citation Ablation Results}
\label{sec:appendix-e2}

The citation-ablation split identifies which artifact types activate citation preference. Rates are percentages over 400 comparisons per variant.

\begin{table*}[!t]
  \centering
  \footnotesize
  \begin{tabular}{>{\raggedright\arraybackslash}p{2.8cm}>{\raggedright\arraybackslash}p{2.5cm}rrrr}
    \toprule
    \textbf{Model} & \textbf{Variant} & \textbf{Tie} & \textbf{Citation win} & \textbf{Base win} & \textbf{Signed effect} \\
    \midrule
    GPT-5.5 & full citation     & 25.2 & 74.5 &  0.2 &  74.2 \\
    GPT-5.5 & source-only       & 41.2 & 58.8 &  0.0 &  58.8 \\
    GPT-5.5 & empty shell       & 52.0 &  0.0 & 48.0 & -48.0 \\
    GPT-5.5 & evidence-only     & 16.2 &  0.0 & 83.8 & -83.8 \\
    GPT-5.5 & sentence-id-only  & 95.5 &  0.0 &  4.5 &  -4.5 \\
    GPT-5.5 & length-control    &  0.5 &  0.0 & 99.5 & -99.5 \\
    Claude Sonnet 4.6 & full citation    & 48.5 & 51.5 &  0.0 &  51.5 \\
    Claude Sonnet 4.6 & source-only      & 71.8 & 28.2 &  0.0 &  28.2 \\
    Claude Sonnet 4.6 & empty shell      & 98.0 &  0.0 &  2.0 &  -2.0 \\
    Claude Sonnet 4.6 & evidence-only    & 98.8 &  0.0 &  1.2 &  -1.2 \\
    Claude Sonnet 4.6 & sentence-id-only & 100.0 &  0.0 &  0.0 &   0.0 \\
    Claude Sonnet 4.6 & length-control   &  0.2 &  0.0 & 99.8 & -99.8 \\
    DeepSeek & full citation      & 53.2 & 46.8 &  0.0 &  46.8 \\
    DeepSeek & source-only        & 90.5 &  9.5 &  0.0 &   9.5 \\
    DeepSeek & empty shell        & 87.2 & 12.2 &  0.5 &  11.8 \\
    DeepSeek & evidence-only      & 83.5 &  0.0 & 16.5 & -16.5 \\
    DeepSeek & sentence-id-only   & 98.8 &  1.0 &  0.2 &   0.8 \\
    DeepSeek & length-control     &  7.8 &  0.0 & 92.2 & -92.2 \\
    \bottomrule
  \end{tabular}
  \captionsetup{skip=3pt}
  \caption{Citation ablation results.}
  \label{tab:ablation-results}
\end{table*}

The strongest positive effects are full-citation and source-only variants. Length-control does not explain the effect.

\subsection{Moderate Prompt Frontier (Full Ladder)}
\label{sec:appendix-e3}

The main paper reports L0 and L4 endpoints in compact form. The full prompt ladder shows how better-plain resolution changes across stricter operating points; the pattern is model-specific and not assumed to be monotonic.

\begin{table*}[!t]
  \centering
  \small
  \begin{tabular}{lrrrrrrrrrr}
    \toprule
    \textbf{Judge} & \multicolumn{2}{c}{\textbf{L0}} & \multicolumn{2}{c}{\textbf{L1}} & \multicolumn{2}{c}{\textbf{L2}} & \multicolumn{2}{c}{\textbf{L3}} & \multicolumn{2}{c}{\textbf{L4}} \\
    \cmidrule(lr){2-3} \cmidrule(lr){4-5} \cmidrule(lr){6-7} \cmidrule(lr){8-9} \cmidrule(lr){10-11}
    & \textbf{better} & \textbf{Tie} & \textbf{better} & \textbf{Tie} & \textbf{better} & \textbf{Tie} & \textbf{better} & \textbf{Tie} & \textbf{better} & \textbf{Tie} \\
    \midrule
    GPT-5.5            & 100.0 & 0.0 & 100.0 & 0.0 & 99.0 & 1.0 & 99.0 & 1.0 &  2.0 & 98.0 \\
    Claude Sonnet 4.6  & 100.0 & 0.0 & 97.5  & 2.5 & 91.0 & 9.0 & 71.5 & 28.5 & 11.0 & 89.0 \\
    DeepSeek           & 99.5  & 0.5 & 96.5  & 3.0 & 82.5 & 17.0 & 79.0 & 21.0 & 15.0 & 85.0 \\
    Qwen2.5-14B-Instruct-AWQ & 99.5 & 0.0 & 97.5 & 2.0 & 89.5 & 10.0 & 90.0 & 8.0 & 25.5 & 74.5 \\
    Gemma-3-12B-IT     & 100.0 & 0.0 & 99.5  & 0.5 & 97.0 & 3.0 & 43.0 & 57.0 & 52.0 & 48.0 \\
    \bottomrule
  \end{tabular}
  \captionsetup{skip=3pt}
  \caption{Moderate prompt frontier (full ladder).}
  \label{tab:moderate-frontier}
\end{table*}

The table reports better-plain and Tie rates; the omitted residual is the worse-cited win rate. Worse-cited rates are near zero across the moderate ladder and never drive the frontier. For example, Qwen L0 is 99.5\% better-plain, 0.5\% worse-cited, and 0.0\% Tie; the largest omitted residual in this compact table is 2.0\%, from Qwen at L3. The frontier is therefore driven primarily by better-plain decisions being converted into Tie, not by renewed preference for the weaker cited candidate.

\subsection{TRACE Recovery on the Moderate Tier}
\label{sec:appendix-e4}

Table~\ref{tab:trace-recovery} reports the TRACE recovery rates on the moderate tier.

\begin{table*}[!t]
  \centering
  \small
  \begin{tabular}{lrrr}
    \toprule
    \textbf{Judge} & \textbf{TRACE better-plain} & \textbf{TRACE worse-cited} & \textbf{TRACE Tie} \\
    \midrule
    GPT-5.5 & 99.0 & 0.0 & 1.0 \\
    Claude Sonnet 4.6 & 100.0 & 0.0 & 0.0 \\
    DeepSeek & 97.0 & 0.0 & 3.0 \\
    Qwen2.5-14B-Instruct-AWQ & 98.5 & 0.0 & 1.5 \\
    Gemma-3-12B-IT & 96.5 & 0.0 & 3.5 \\
    \bottomrule
  \end{tabular}
  \captionsetup{skip=3pt}
  \caption{TRACE recovery on the moderate tier.}
  \label{tab:trace-recovery}
\end{table*}

TRACE recovery supports the interpretation that resolution loss is protocol-dependent. Removing citation-like display is not sufficient; the normalization step must preserve evidence-chain quality.

\subsection{FinQA Moderate Robustness Check}
\label{sec:appendix-e4b}

The 50-pair FinQA construction is run in both candidate orders. Table~\ref{tab:finqa-moderate} reports the ranges available across L0--L2 and the key strict and TRACE endpoints.

\begin{table*}[!t]
  \centering
  \small
  \begin{tabular}{lrrrr}
    \toprule
    \textbf{Judge} & \textbf{L0--L2 better-plain} & \textbf{L4 Tie} & \textbf{TRACE better-plain} & \textbf{TRACE Tie} \\
    \midrule
    GPT-5.5 & 95--97 & 91 & 98 & 0 \\
    Claude Sonnet 4.6 & 92--98 & 100 & 100 & 0 \\
    DeepSeek & 88--95 & 99 & 100 & 0 \\
    \bottomrule
  \end{tabular}
  \captionsetup{skip=3pt}
  \caption{FinQA moderate-gap robustness results (\%). Each condition contains 100 ordered comparisons.}
  \label{tab:finqa-moderate}
\end{table*}

The DeepSeek L3 better-plain rate is 46\%, demonstrating loss before the L4 endpoint. On the HotpotQA tier, exact McNemar tests show significant L0-to-L4 losses and L4-to-TRACE recoveries for every commercial judge ($p<0.001$). Claude's L0-to-L4 better-plain drop is 89.0 percentage points (95\% paired-bootstrap CI [84.5, 93.0]); GPT-5.5's drop is 98.0 points (CI [96.0, 99.5]). Bootstrap intervals use 2,000 resamples over pair-by-order units. The L3-to-TRACE comparison is significant for Claude and DeepSeek; it is not significant for GPT-5.5 ($p=1.0$), consistent with GPT-5.5's near-absence of L3 loss.

\subsection{Equivalence Suppression by Split}
\label{sec:appendix-e5}

Main-paper Figure~\ref{fig:equivalence} (equivalence suppression) condenses these numbers into per-judge bars. See Table~\ref{tab:equivalence-suppression} for the full breakdown.

\begin{table*}[!t]
  \centering
  \small
  \begin{tabular}{>{\raggedright\arraybackslash}p{2.8cm}>{\raggedright\arraybackslash}p{3cm}rr}
    \toprule
    \textbf{Judge} & \textbf{Split} & \textbf{Plain cited win} & \textbf{Strict Tie} \\
    \midrule
    GPT-5.5 & HotpotQA mechanical      & 85.8 & 100.0 \\
    GPT-5.5 & FinQA mechanical         & 83.5 & 100.0 \\
    GPT-5.5 & Natural citation-looking & 93.5 &  99.9 \\
    Claude Sonnet 4.6 & HotpotQA mechanical      & 50.9 & 100.0 \\
    Claude Sonnet 4.6 & FinQA mechanical         & 42.3 & 100.0 \\
    Claude Sonnet 4.6 & Natural citation-looking & 22.8 & 100.0 \\
    DeepSeek & HotpotQA mechanical      & 47.2 & 100.0 \\
    DeepSeek & FinQA mechanical         & 44.2 & 100.0 \\
    DeepSeek & Natural citation-looking & 24.6 &  99.8 \\
    \bottomrule
  \end{tabular}
  \captionsetup{skip=3pt}
  \caption{Equivalence suppression under strict prompting.}
  \label{tab:equivalence-suppression}
\end{table*}

These results motivate the first half of the paper: strict anti-citation prompting can indeed suppress presentation preference in equivalence settings.

\subsection{Correctness Retention}
\label{sec:appendix-e6}

Clear and reverse correctness conflicts show that resolution loss is not simply correctness collapse. The decoupled correctness-retention probe remains at or above 93.0\% correct-base wins across commercial judges, supporting the claim that the main failure mode lies in moderate support-chain gaps rather than obvious correct-versus-wrong cases. L4 strict-prompt correctness-family results are included in the supplementary artifact; Table~\ref{tab:correctness-retention} is a targeted decoupled probe and should not be read as the L4 strict-prompt endpoint.

The decoupled correctness-retention probe rates appear in Table~\ref{tab:correctness-retention}.

\begin{table*}[!t]
  \centering
  \small
  \begin{tabular}{lrrrr}
    \toprule
    \textbf{Model} & \textbf{$n$} & \textbf{Correct base win} & \textbf{Wrong cited win} & \textbf{Tie} \\
    \midrule
    GPT-5.5 & 300 & 97.7 & 1.0 & 1.3 \\
    Claude Sonnet 4.6 & 300 & 98.0 & 1.0 & 1.0 \\
    DeepSeek & 300 & 93.0 & 2.0 & 5.0 \\
    \bottomrule
  \end{tabular}
  \caption{Decoupled correctness-retention probe.}
  \label{tab:correctness-retention}
\end{table*}

For full plain/strict correctness-family reporting, the supplementary artifact includes per-judge, per-prompt-family, per-paraphrase, and per-order rates.

\subsection{Position Controls}
\label{sec:appendix-e7}

Position-only controls use \emph{identical} plain/no-citation candidates in both orders. Any non-Tie decision in this setting would indicate generic position bias. Table~\ref{tab:position-controls} shows that the controls return all Tie.

\begin{table*}[!t]
  \centering
  \small
  \begin{tabular}{lrrrr}
    \toprule
    \textbf{Model} & \textbf{$n$} & \textbf{First-position win} & \textbf{Second-position win} & \textbf{Tie} \\
    \midrule
    Claude Sonnet 4.6 & 1200 & 0.0 & 0.0 & 100.0 \\
    DeepSeek & 400 & 0.0 & 0.0 & 100.0 \\
    \bottomrule
  \end{tabular}
  \caption{Position-only controls.}
  \label{tab:position-controls}
\end{table*}

\textbf{Why this does not contradict the citation-order gaps reported in \S5.1.} The 100\% Tie above holds when \emph{both} candidates are identical plain text. The citation-order gaps in the main paper arise on pairs where one candidate carries citation-like display and the other does not, presented in different orders. Position-only controls therefore confirm that the order coupling is \emph{citation-conditioned}, not generic---i.e., position bias does not emerge until citation-like display is introduced.

GPT-5.5 is not included in this targeted position-only control table because the added controls were run for judges that showed the largest citation-conditioned order gaps in the commercial audit. In the mechanical plain split, GPT-5.5's cited-win rate differs by 2.7 percentage points across candidate orders, whereas DeepSeek and Claude required additional controls to rule out generic position-only behavior. The absence of a GPT-5.5 row should therefore be read as scoped control coverage, not as a claim that GPT-5.5 has no order sensitivity in general.

\subsection{Workflow-Ranking Probe}
\label{sec:appendix-e8}

The ranking probe uses four workflow variants---\texttt{answer\_only}, \texttt{plain\_full}, \texttt{cited\_full}, and \texttt{cited\_partial}. It contains 200 HotpotQA queries, all six unordered system pairs per query, and both candidate orders, yielding 2,400 judge rows per protocol/model setting (each system appears in 1,200 ordered comparisons). Scores count one point for a win and one-half point for a Tie, normalized to percentages.

\begin{table*}[!t]
  \centering
  \small
  \begin{tabular}{>{\raggedright\arraybackslash}p{2.8cm}>{\raggedright\arraybackslash}p{2.5cm}rrrr}
    \toprule
    \textbf{Judge} & \textbf{Protocol} & \textbf{answer\_only} & \textbf{plain\_full} & \textbf{cited\_full} & \textbf{cited\_partial} \\
    \midrule
    GPT-5.5 & plain preference & 19.38 & 66.62 & 98.17 & 15.83 \\
    GPT-5.5 & TRACE            &  2.62 & 83.00 & 83.17 & 31.21 \\
    Claude Sonnet 4.6 & plain preference &  2.54 & 76.83 & 89.29 & 31.33 \\
    Claude Sonnet 4.6 & TRACE            &  2.38 & 83.12 & 83.12 & 31.37 \\
    DeepSeek & plain preference &  1.71 & 80.12 & 82.50 & 35.67 \\
    DeepSeek & TRACE            &  2.04 & 80.33 & 82.75 & 34.88 \\
    \bottomrule
  \end{tabular}
  \caption{Workflow-ranking probe system scores.}
  \label{tab:workflow-scores}
\end{table*}

\subsection{Direct Workflow-Pair Comparisons}
\label{sec:appendix-e8b}

Direct full-workflow comparisons show the presentation effect more explicitly:

\begin{table*}[!t]
  \centering
  \small
  \begin{tabular}{>{\raggedright\arraybackslash}p{2.8cm}>{\raggedright\arraybackslash}p{2.5cm}rrr}
    \toprule
    \textbf{Judge} & \textbf{Protocol} & \textbf{\texttt{plain\_full} wins} & \textbf{\texttt{cited\_full} wins} & \textbf{Tie} \\
    \midrule
    GPT-5.5 & plain preference  &   0 & 371 &  29 \\
    GPT-5.5 & correctness-first &   0 &   2 & 398 \\
    GPT-5.5 & TRACE             &   0 &   0 & 400 \\
    Claude Sonnet 4.6 & plain preference  &   0 & 147 & 253 \\
    Claude Sonnet 4.6 & correctness-first &   0 &   0 & 400 \\
    Claude Sonnet 4.6 & TRACE             &   0 &   1 & 399 \\
    DeepSeek & plain preference  &   1 &   4 & 395 \\
    DeepSeek & correctness-first &   0 &   0 & 400 \\
    DeepSeek & TRACE             &   5 &  23 & 372 \\
    \bottomrule
  \end{tabular}
  \caption{Direct \texttt{plain\_full} versus \texttt{cited\_full} comparisons.}
  \label{tab:direct-comparisons}
\end{table*}

The direct-comparison table uses 400 ordered rows per judge/protocol, corresponding to 200 queries in both candidate orders. It should therefore not be compared directly to order-collapsed 200-row summaries.

\section{Soft Tier: Calibrated Tie Counterpart}
\label{sec:appendix-f}

Main-paper \S5.3 argues that Tie is calibrated on soft-boundary pairs but resolution-destroying on moderate-gap pairs. The soft-tier prompt ladder, reported here, supports the calibrated half of that claim.

\begin{table*}[!t]
  \centering
  \small
  \begin{tabular}{lrrrrrrrrrr}
    \toprule
    \textbf{Judge} & \multicolumn{2}{c}{\textbf{L0}} & \multicolumn{2}{c}{\textbf{L1}} & \multicolumn{2}{c}{\textbf{L2}} & \multicolumn{2}{c}{\textbf{L3}} & \multicolumn{2}{c}{\textbf{L4}} \\
    \cmidrule(lr){2-3} \cmidrule(lr){4-5} \cmidrule(lr){6-7} \cmidrule(lr){8-9} \cmidrule(lr){10-11}
    & \textbf{worse} & \textbf{Tie} & \textbf{worse} & \textbf{Tie} & \textbf{worse} & \textbf{Tie} & \textbf{worse} & \textbf{Tie} & \textbf{worse} & \textbf{Tie} \\
    \midrule
    GPT-5.5           & 50.5 & 32.2 & 0.0 & 78.5 & 0.0 & 99.5 & 0.0 & 99.0 & 0.0 & 100.0 \\
    Claude Sonnet 4.6 & 10.0 & 51.5 & 0.0 & 80.5 & 0.0 & 93.0 & 0.0 & 98.5 & 0.0 & 100.0 \\
    DeepSeek          & 24.5 & 70.0 & 0.0 & 97.5 & 0.0 & 98.0 & 0.0 & 99.0 & 0.0 & 99.5 \\
    \bottomrule
  \end{tabular}
  \caption{Soft-tier prompt frontier. ``worse'' is the worse-cited win
  rate, ``Tie'' is the Tie rate; the omitted residual is the
  better-plain rate, which is small throughout because the soft tier
  is intentionally constructed near the human decision boundary.}
  \label{tab:soft-frontier}
\end{table*}

\begin{figure}[H]
  \centering
  \includegraphics[width=\columnwidth]{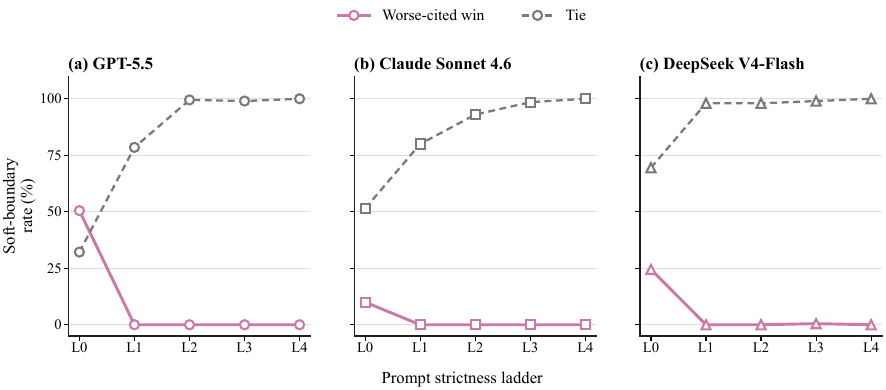}
  \caption{Soft-tier prompt-ladder behavior}
  \label{fig:soft-tier}
\end{figure}

The contrast between Table~\ref{tab:soft-frontier} (soft) and
Table~\ref{tab:moderate-frontier} (moderate) is the core empirical
evidence behind the three-meanings-of-Tie distinction in \S5.3: when
humans agree there is a direction, judges that abstain are losing
resolution; when humans themselves do not commit to a direction,
judges that abstain are calibrated.

\section{Rationale Keyword Analysis}
\label{sec:appendix-g}

This section supports the main paper's claim (\S5.1) that judges respond to source-like authority or traceability display rather than to additional tokens alone. Rationale analysis is used only as supporting evidence; it does not reveal hidden mechanisms and is not treated as causal proof.

We analyze public judge rationales for source/citation-related keywords and compare citation-win rationales with Tie rationales.

\begin{table*}[!t]
  \centering
  \small
  \begin{tabular}{>{\raggedright\arraybackslash}p{2.8cm}>{\raggedright\arraybackslash}p{3cm}lr}
    \toprule
    \textbf{Judge} & \textbf{Split} & \textbf{Decision type} & \textbf{Keyword rate} \\
    \midrule
    GPT-5.5 & HotpotQA mechanical & citation win & 100.0 \\
    GPT-5.5 & FinQA mechanical    & citation win & 100.0 \\
    GPT-5.5 & HotpotQA mechanical & Tie          &  96.5 \\
    Claude Sonnet 4.6 & HotpotQA mechanical      & citation win & 100.0 \\
    Claude Sonnet 4.6 & FinQA mechanical         & citation win & 100.0 \\
    Claude Sonnet 4.6 & HotpotQA mechanical      & Tie          &  40.1 \\
    Claude Sonnet 4.6 & FinQA mechanical         & Tie          &  30.9 \\
    Claude Sonnet 4.6 & Natural citation-looking & Tie          &  97.1 \\
    DeepSeek & HotpotQA mechanical & citation win & 100.0 \\
    DeepSeek & HotpotQA mechanical & Tie          &  15.6 \\
    \bottomrule
  \end{tabular}
  \caption{Source/citation keyword rates in judge rationales (\%).}
  \label{tab:keyword-rates}
\end{table*}

Citation-win rationales consistently describe cited outputs as more supported, traceable, or verifiable. Tie rationales can still mention citation-like display, especially for GPT-5.5 and natural-pair Claude results. We interpret this as a distinction between \emph{cue perception} and \emph{cue use}: a judge may notice citation-like display while declining to convert it into a preference.

Table~\ref{tab:keyword-rates} reports the rationale subsets used for the keyword audit rather than a complete rationale matrix over every split and model. Missing cells indicate that the corresponding rationale subset was not part of the released keyword-rate audit or was too small for stable reporting.

\section{Reproducibility Checklist}
\label{sec:appendix-h}

Released artifacts include:

\begin{itemize}
  \item \texttt{splits/}: benchmark split files with stable item identifiers, source dataset tags, target relation labels, candidate-order fields, and artifact-side metadata;
  \item \texttt{prompts/}: full L0--L4 prompt templates, prompt paraphrases, TRACE normalization prompts, and masked judging prompts;
  \item \texttt{raw\_outputs/}: raw A/B/Tie outputs, rationales, parse status, retry status, and unparsed responses;
  \item \texttt{metadata/}: model identifiers, endpoint aliases, endpoint-returned snapshots when available, call dates, inference parameters, and provider response metadata;
  \item \texttt{human\_validation/}: de-identified annotation workbooks, guidelines, blind candidate order, per-item labels, and validation summaries;
  \item \texttt{scripts/}: parsers and analysis scripts for win rates, Tie rates, signed effects, order gaps, PABAK, Gwet's AC1, Wilson intervals, and bootstrap confidence intervals;
  \item \texttt{requirements.txt}: pinned non-standard Python runtime dependency for annotation-workbook processing; the remaining scripts use the Python standard library;
  \item \texttt{figures/}: scripts and source tables used to generate the main-paper figures and appendix tables;
  \item \texttt{logs/}: amendment logs for any model-name, endpoint-name, parsing, or data-cleaning changes.
\end{itemize}

The artifact is de-identified. It does not include private API keys, account identifiers, or personally identifying annotator information.

\textbf{Caveat on endpoint version pinning.} Some commercial endpoints expose limited version information. For DeepSeek the endpoint-returned alias (\texttt{deepseek-v4-flash}) does not always reveal the deeper provider-side snapshot; for Claude Sonnet 4.6, the snapshot is not exposed by the gateway. Released logs therefore preserve all available metadata returned by the endpoint and report the call-date window, but exact provider-side snapshots may not be recoverable. We treat this as a reproducibility limitation rather than a defect of the audit design.

\textbf{Prompt paraphrase coverage.} The prompt ladder uses one paraphrase per level for soft and most HotpotQA moderate analyses, with three paraphrases reserved for clear and reverse correctness conflicts. The FinQA check adds two faithful L2 paraphrases, one faithful L3 paraphrase, and one separately labeled L4-like boundary variant. This remains a targeted audit rather than an exhaustive prompt search; the supplementary artifact lists all tested prompts.

\end{document}